\documentclass[runningheads]{llncs}
\usepackage[T1]{fontenc}
\usepackage{graphicx}
\usepackage{listings}
\usepackage{formaljitterbug}
\usepackage{cleveref}
\usepackage{tikz} 
\usepackage{tikz-cd}
\usepackage{caption}
\usetikzlibrary{positioning}

\begin{document}
\title{Well-Behaved (Co)algebraic Semantics of Regular Expressions in Dafny}

\author{Stefan Zetzsche\inst{1} \and
Wojciech R{\' o}{\. z}owski\inst{2}}

\authorrunning{S. Zetzsche and W. R{\' o}{\. z}owski}

\institute{Amazon Web Services, United Kingdom \\
\email{stefanze@amazon.co.uk} \and
University College London, United Kingdom \\
\email{w.rozowski@cs.ucl.ac.uk	}
}

\maketitle        

\begin{abstract}
Regular expressions are commonly understood in terms of their denotational semantics, that is, through formal languages -- the regular languages. This view is inductive in nature: two primitives are equivalent if they are constructed in the same way. Alternatively, regular expressions can be understood in terms of their operational semantics, that is, through deterministic finite automata. This view is coinductive in nature: two primitives are equivalent if they are deconstructed in the same way. It is implied by Kleene's famous theorem that both views are equivalent: regular languages are precisely the formal languages accepted by deterministic finite automata. In this paper, we use Dafny, a verification-aware programming language, to formally verify, for the first time, what has been previously established only through proofs-by-hand: the two semantics of regular expressions are well-behaved, in the sense that they are in fact one and the same, up to pointwise bisimilarity. At each step of our formalisation, we propose an interpretation in the language of Coalgebra. We found that Dafny is particularly well suited for the task due to its inductive and coinductive features and hope our approach serves as a blueprint for future generalisations to other theories.

\end{abstract}

\section{Introduction}

\label{sec:introduction}

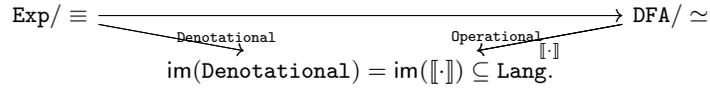
\begin{figure*}[t]
\center
		\begin{tikzcd}[row sep=0.2cm]
			\texttt{Exp}/\equiv \arrow{rr}[below]{} \arrow{rr}[above]{} \arrow{dr}[right]{\texttt{Denotational}} \arrow{dr}[above]{}&& \texttt{DFA}/\simeq \arrow{dl}[above]{} \arrow{dl}[below]{\llbracket \cdot \rrbracket} \arrow{dl}[left]{\texttt{Operational}} \\
			    & \textsf{im}(\texttt{Denotational}) = \textsf{im}(\llbracket \cdot \rrbracket) \subseteq \texttt{Lang}.
		\end{tikzcd}	
	\caption{The triptych of regular expressions, deterministic finite automata, and regular languages.}
	\label{fig:triptichon}
\end{figure*}

Regular expressions stand as one of the most ubiquitous formalisms in all of theoretical computer science. Their inception can be traced back all the way to Kleene's seminal paper in 1951 \cite{kleene1951representation}. Today, they play a pivotal role as a foundational element for a wide spectrum of applications~\cite{Friedl:2006:Mastering,Owens:2009:Regular,Ausaf:2016:POSIX}, encompassing text searching, pattern matching, lexical analysis, and more.

Typically, regular expressions are understood denotationally, through the formal languages, that is, sets of finite words over a fixed alphabet, that they denote. This view is inductive in nature, in the sense that the denotational semantics of regular expressions is constructed from the bottom-up by following the finite inductive structure of an expression.

Alternatively, regular expressions can be understood operationally, through the lenses of deterministic finite automata. This view is coinductive in nature, in the sense that the operational semantics of regular expressions is assigned from the top-down, by deconstructing an expression, following the coinductive nature of a potentially infinite language initially observed by Brzozowski \cite{brzozowski1964derivatives}. 

 One of Kleene's many contributions was to show that the denotational and operational semantics of regular expressions are two sides of the same coin -- they are \emph{well-behaved}. That is, the denotational interpretation of a regular expression matches exactly the observable behaviour of its operational interpretation. Kleene's theorem is of great practical significance, too: for any given regular expression that represents a text pattern, one can derive, in a canonical way, an automaton that gives rise to a deterministic algorithm that decides, in finite time, whether some given string matches the text pattern specified by the expression~\cite{Hopcroft:1971:Linear,Holzer:2011:Complexity}.  The full triptych of regular expressions, deterministic finite automata, and regular languages is depicted in \Cref{fig:triptichon}: the set of regular expressions modulo the axioms of \emph{Kleene Algebra} \cite{Kozen:1994:Completeness} ($\equiv$), the set of deterministic finite automata modulo behavioural equivalence ($\simeq$), and the set of regular languages are in bijection. The composition on the right-hand side of the diagram can be seen as a function that assigns the operational semantics to an expression.
 
 In more recent years, a more general approach to automata through the lenses of category theory has become popular: state-based systems are generalised as \emph{coalgebras} over an endofunctor \cite{Rutten:00:Universal,Gumm:2000:Elements,jacobs2012trace}. There are many advantages to using the coalgebraic abstraction of state-based systems. Among others, it allows one to set aside irrelevant specifics of concrete instantiations, and instead work with elegant, universal properties. Of particular interest for us are systems that have both an algebraic (inductive)
and a coalgebraic (coinductive) component. 
 
 In this paper, we use the built-in inductive and coinductive reasoning capabilities of \emph{Dafny} \cite{dafny-website}, a programming language and static verifier, to formalise the denotational and operational semantics of regular expressions and formally prove that they are well-behaved, that is, coincide pointwise, up to bisimilarity\footnote{The full Dafny source code is available at \cite{Zetzsche:2024:Well-behaved}.}. 
Dafny is a statically typed programming language with native support for writing and verifying specifications about programs that was first developed by Leino at Microsoft Research \cite{microsoft-research,leino2010dafny}. Dafny combines various paradigms such as imperative, functional, and object-oriented programming and supports common programming concepts such as inductive datatypes, immutable and mutable data structures, lambda functions, and subset types. Dafny can be integrated with common software development IDEs such as VSCode and emacs. It has been used in academia for research and the teaching of program verification \cite{noble2022more}, as well as in industry by e.g. Amazon \cite{aws-encryption-sdk}, ConsenSys \cite{cassez2023formal}, and Intel \cite{yang2023towards}. Teaching material is available online \cite{leino2poweruser,awsteaching} and in print \cite{leino2023program}. A blog covers various aspects of the Dafny ecosystem \cite{dafny-blog}.

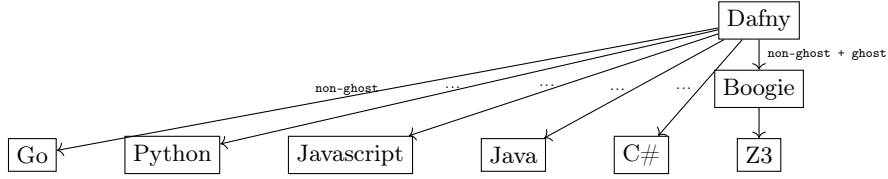
\begin{figure*}[t]
\centering
		\begin{tikzpicture}[align=center, node distance=0.4cm and 0.9cm]
		\node[draw] (1) {Dafny};
		\node[draw]  (2) [below= of 1] {Boogie};
		\node[draw]  (3) [below= of 2] {Z3};
		\node[draw]  (5) [left= of 3]{C\#};
		\node[draw] (6) [left= of 5] {Java};
		\node[draw] (7) [left= of 6] {Javascript};
		\node[draw] (8) [left= of 7] {Python};
		\node[draw] (9) [left= of 8] {Go};
		
		\draw[->] (1) edge node[right] {\tiny \texttt{non-ghost + ghost}} (2);
		\draw[->] (2) -- (3);
		\draw[->] (1) edge node[left] {\tiny ...} (5);
				\draw[->] (1) edge node[left] {\tiny ...} (6);
				\draw[->] (1) edge node[left] {\tiny ...} (7);
				\draw[->] (1) edge node[left] {\tiny ...} (8);
				\draw[->] (1) edge node[left] {\tiny \texttt{non-ghost}} (9);

	\end{tikzpicture}
\caption{The compilation pipeline of Dafny.}
\label{fig:compilation_of_dafny}
\end{figure*}

One of the features of Dafny is that it allows the clear distinction between an idealised mathematical specification and an efficient implementation thereof. As a first example, consider the following purely functional specification of the Fibonacci sequence:

\begin{lstlisting}[language=dafny]
function Fib(n: nat): nat {
  if n <= 1 then n else Fib(n - 1) + Fib(n - 2)
}
\end{lstlisting}
While elegant in its recursive definition, \lstinline[language=dafny, basicstyle=\small\ttfamily]{Fib} is not particularly efficient. A more realistic implementation is given by the imperative method \lstinline[language=dafny, basicstyle=\small\ttfamily]{ComputeFib} below:
\begin{lstlisting}[language=dafny]
method ComputeFib(n: nat) returns (b: nat)
  ensures b == Fib(n) 
{
  var c := 1;
  b := 0;
  for i := 0 to n
    invariant b == Fib(i) && c == Fib(i + 1)
  {
    b, c := c, b + c;
  }
}
\end{lstlisting}

Dafny allows us to extend the method signature with an \lstinline[language=dafny, basicstyle=\small\ttfamily]{ensures} clause, which in this case indicates that the outputs of \lstinline[language=dafny, basicstyle=\small\ttfamily]{Fib} and \lstinline[language=dafny, basicstyle=\small\ttfamily]{ComputeFib} coincide on all possible inputs. To aid Dafny with proving the correctness of the \lstinline[language=dafny, basicstyle=\small\ttfamily]{ensures} clause, we have to identify an appropriate \lstinline[language=dafny, basicstyle=\small\ttfamily]{invariant} of the \lstinline[language=dafny, basicstyle=\small\ttfamily]{for} loop in the body of \lstinline[language=dafny, basicstyle=\small\ttfamily]{ComputeFib}. While not displayed, \lstinline[language=dafny, basicstyle=\small\ttfamily]{ensures} admits a dual, the \lstinline[language=dafny, basicstyle=\small\ttfamily]{requires} clause, which is used to restrict the domain of functions and methods to a subset. The two clauses are best thought of in terms of pre- and postconditions in the spirit of Hoare triples~\cite{Hoare:1969:Axiomatic}.

As illustrated above, Dafny programs contain both so-called \emph{ghost} and \emph{non-ghost} parts. 
Ghost code is meant for the specification of the behaviour of functions and the proof thereof only, not for compilation. 
 Functions, methods, and variables can be marked ghost with a designated keyword. A method that is ghost and doesn't modify the heap is called a lemma. Pre- and postconditions, assertions, and loop invariants are always considered ghost.
The Dafny verifier translates the ghost and non-ghost parts of a Dafny program into a program in the intermediate verification language Boogie \cite{barnett2006boogie}, such that the correctness of the output program implies the correctness of the input program. To verify the correctness of a Boogie program, a verification condition is generated from it and passed to the SMT solver Z3 \cite{de2008z3} (\Cref{fig:compilation_of_dafny}). Besides Dafny, there are other verification-aware languages built on top of Boogie and Z3 (e.g. VCC \cite{cohen2009vcc} and Spec\# \cite{barnett2011specification}). The non-ghost part of a verified Dafny program can be compiled to C\#, Java, Javascript, Python, and Go, making possible the integration of verified code with an existing code base (\Cref{fig:compilation_of_dafny}).

We found that Dafny is particularly well suited for the task and would like our approach to serve as a blueprint for future generalisations to other theories. At each step of our formalisation, we propose an interpretation in the language of Coalgebra. We hope that the presentation is accessible both for readers that are familiar with Coalgebra but not so much with Dafny, and for readers unexposed to Coalgebra, but experienced in Dafny.

In detail, the paper makes the following contributions:

\begin{itemize}
	\item We formalise regular expressions as an inductive datatype (\Cref{sec:regular_expressions_as_datatype}) and formal languages as a coinductive codatatype (\Cref{sec:formal_languages_as_codatatype}). We introduce the concept of bisimilarity of languages (\Cref{sec:bisimilarity_and_coinduction}). We equip languages with an algebraic structure (\Cref{sec:an_algebra_of_formal_languages}) and in consequence define the denotational semantics of regular expressions as an induced function from regular expressions to formal languages (\Cref{sec:denotational_semantics_as_induced_morphism}). Finally, we prove that the latter preserves algebraic structures up to pointwise bisimilarity (\Cref{sec:denotational_semantics_as_algebra_homomorphism}). At each step, we propose an interpretation in the language of Coalgebra.
	\item We equip the set of regular expressions with a coalgebraic structure of the type of unpointed deterministic automata (\Cref{sec:a_coalgebra_of_regular_expressions}). We then formalise the operational semantics of regular expressions as an induced function from regular expressions to formal languages (\Cref{sec:operational_semantics_as_induced_morphism}). Finally, we prove that the latter preserves coalgebraic structures (\Cref{sec:operational_semantics_as_coalgebra_homomorphism}).
	\item We show that the function that formalises the denotational semantics is also a coalgebra homomorphism (\Cref{sec:denotational_semantics_as_coalgebra_homomorphism}), and that coalgebra homomorphisms are unique up to pointwise bisimilarity (\Cref{sec:coalgebra_homomorphisms_are_unique}). We deduce that the denotational and operational semantics coincide, up to pointwise bisimilarity (\Cref{sec:denotational_and_operational_semantics_are_bisimilar}). Finally, we show that the function that formalises the operational semantics is also an algebra homomorphism (\Cref{sec:operational_semantics_as_algebra_homomorphism}).
\end{itemize}

\section{Denotational Semantics}

\label{sec:denotational_semantics}

In this section, we define, in Dafny, regular expressions and formal languages, introduce the concept of bisimilarity, formalise
the \emph{denotational} semantics of regular expressions as a function from regular expressions to formal languages, and
prove that the latter is an algebra homomorphism.

\subsection{Regular Expressions as Datatype}

\label{sec:regular_expressions_as_datatype}

We define the set of regular expressions parametric in an alphabet \lstinline[language=dafny, basicstyle=\small\ttfamily]{A} as an inductive datatype:

\begin{lstlisting}[language=dafny]
datatype Exp<A> = | Zero | One | Char(A) | Plus(Exp, Exp) | Comp(Exp, Exp) | Star(Exp)
\end{lstlisting}

Note that above, and later, we make use of Dafny's type parameter completion \cite{Leino:typeparametercompletion}, which allows us to write \lstinline[language=dafny, basicstyle=\small\ttfamily]{Exp} instead of \lstinline[language=dafny, basicstyle=\small\ttfamily]{Exp<A>}.

The definition above captures that a regular expression is either a primitive character \lstinline[language=dafny, basicstyle=\small\ttfamily]{Char(a)}, a non-deterministic
choice between two regular expressions \lstinline[language=dafny, basicstyle=\small\ttfamily]{Plus(e1, e2)}, a sequential composition of two regular expressions
\lstinline[language=dafny, basicstyle=\small\ttfamily]{Comp(e1, e2)}, a finite number of self-iterations \lstinline[language=dafny, basicstyle=\small\ttfamily]{Star(e)}, or one of the constants \lstinline[language=dafny, basicstyle=\small\ttfamily]{Zero} (the unit of \lstinline[language=dafny, basicstyle=\small\ttfamily]{Plus})
and \lstinline[language=dafny, basicstyle=\small\ttfamily]{One} (the unit of \lstinline[language=dafny, basicstyle=\small\ttfamily]{Comp}). At a higher level, the above defines \lstinline[language=dafny, basicstyle=\small\ttfamily]{Exp<A>} as the smallest algebraic structure
that is equipped with two constants, contains all elements of type \lstinline[language=dafny, basicstyle=\small\ttfamily]{A}, and is closed under two binary operations
and one unary operation. Even more abstractly, \lstinline[language=dafny, basicstyle=\small\ttfamily]{Exp<A>} can be viewed as the initial algebra for the set endofunctor $\Sigma$ defined on objects by $\Sigma X = 1 + 1 + A + X^2 + X^2 + X$.

\subsection{Formal Languages as Codatatype}

\label{sec:formal_languages_as_codatatype}

We define the set of formal languages parametric in an alphabet \lstinline[language=dafny, basicstyle=\small\ttfamily]{A} as a coinductive codatatype:

\begin{lstlisting}[language=dafny]
codatatype Lang<!A> = Alpha(eps: bool, delta: A -> Lang<A>)
 \end{lstlisting} 
\vspace{-0.3cm}

 Note that we used Dafny's type-parameter mode \lstinline[language=dafny, basicstyle=\small\ttfamily]{!}, which indicates that there could be strictly more values of type
\lstinline[language=dafny, basicstyle=\small\ttfamily]{Lang<A>} than values of type \lstinline[language=dafny, basicstyle=\small\ttfamily]{A}, for any type \lstinline[language=dafny, basicstyle=\small\ttfamily]{A}, and that there is no subtype relation between \lstinline[language=dafny, basicstyle=\small\ttfamily]{Lang<A>} and
\lstinline[language=dafny, basicstyle=\small\ttfamily]{Lang<B>}, for any two types \lstinline[language=dafny, basicstyle=\small\ttfamily]{A}, \lstinline[language=dafny, basicstyle=\small\ttfamily]{B}. A more detailed explanation of the topic is available at \cite{Leino:typeparameter}.

To some, our way of modelling formal languages might seem odd at first sight. Typically, a formal language is defined intrinsically, as a set of finite sequences, that is, an element of type $\mathcal{P}(A^*)$ or
\lstinline[language=dafny, basicstyle=\small\ttfamily]{iset<seq<A>>} in Dafny. In our approach, we instead treat languages extrinsically, in terms of their universal property: it is well known that
\lstinline[language=dafny, basicstyle=\small\ttfamily]{iset<seq<A>>} forms the greatest abstract coalgebraic structure \lstinline[language=dafny, basicstyle=\small\ttfamily]{S} that is equipped with functions \lstinline[language=dafny, basicstyle=\small\ttfamily]{eps: S -> bool} and \lstinline[language=dafny, basicstyle=\small\ttfamily]{delta: S -> (A -> S)}. Indeed, for any set U of
finite sequences, we can verify whether U contains the empty sequence, \lstinline[language=dafny, basicstyle=\small\ttfamily]{U.eps == ([] in U)}, and for any \lstinline[language=dafny, basicstyle=\small\ttfamily]{a: A} we can transition to derivative \lstinline[language=dafny, basicstyle=\small\ttfamily]{U.delta(a) == (iset s | [a] + s in U)}. In the language of Coalgebra, \lstinline[language=dafny, basicstyle=\small\ttfamily]{Lang<!A>} can be modelled as the final coalgebra for the set endofunctor $B$ defined on objects by $BX = 2 \times X^A$ \cite{Rutten:00:Universal}. Coalgebras for the functor $B$ correspond precisely to unpointed deterministic automata, and the final object among them provides a universal semantic domain for their behaviour.

We choose the more abstract perspective on formal languages as it hides irrelevant specifics and thus allows us to write more elegant proofs. With this decision, we follow a coalgebraic characterisation of formal languages in Isabelle \cite{traytel2017formal}, but depart from e.g. previous formalisations in Coq \cite{moreira2015deciding}.

\subsection{An Algebra of Formal Languages}

\label{sec:an_algebra_of_formal_languages}

If one thinks of a formal language as a set of finite sequences, one will soon realise that languages admit
quite a bit of algebraic structure. In fact, it becomes clear that formal languages can be equipped with the same type of algebraic
structure as regular expressions.

First, there exists the empty language \lstinline[language=dafny, basicstyle=\small\ttfamily]{Zero()} that contains no words at all. Under the view above, we find
\lstinline[language=dafny, basicstyle=\small\ttfamily]{Zero().eps == false} since the empty set does not contain the empty
sequence, and \lstinline[language=dafny, basicstyle=\small\ttfamily]{Zero().delta(a) == Zero()}, since the derivative \lstinline[language=dafny, basicstyle=\small\ttfamily]@iset s | [a] + s in iset{}@ with respect to any \lstinline[language=dafny, basicstyle=\small\ttfamily]{a: A} yields again the empty set. We thus define:

\begin{lstlisting}[language=dafny]
function Zero<A>(): Lang {
  Alpha(false, (a: A) => Zero())
}
\end{lstlisting} 

Using similar reasoning, we additionally formalise i) the language
\lstinline[language=dafny, basicstyle=\small\ttfamily]{One()} that contains only the empty sequence; ii) for any \lstinline[language=dafny, basicstyle=\small\ttfamily]{a: A} the language \lstinline[language=dafny, basicstyle=\small\ttfamily]{Singleton(a)} that consists of
only the word \lstinline[language=dafny, basicstyle=\small\ttfamily]{[a]}; iii) the language \lstinline[language=dafny, basicstyle=\small\ttfamily]{Plus(L1, L2)} which consists of the union of the languages \lstinline[language=dafny, basicstyle=\small\ttfamily]{L1} and \lstinline[language=dafny, basicstyle=\small\ttfamily]{L2};
iv) the language \lstinline[language=dafny, basicstyle=\small\ttfamily]{Comp(L1, L2)} that consists of all possible concatenation of words in \lstinline[language=dafny, basicstyle=\small\ttfamily]{L1} and \lstinline[language=dafny, basicstyle=\small\ttfamily]{L2}; and v) the
language \lstinline[language=dafny, basicstyle=\small\ttfamily]{Star(L)} that consists of all finite compositions of \lstinline[language=dafny, basicstyle=\small\ttfamily]{L} with itself. Our definitions match what is well-known as \emph{Brzozowski derivatives} \cite{brzozowski1964derivatives}:

\begin{lstlisting}[language=dafny]
function One<A>(): Lang {
  Alpha(true, (a: A) => Zero())
}

function Singleton<A(==)>(a: A): Lang {
  Alpha(false, (b: A) => if a == b then One() else Zero())
}

function {:abstemious} Plus<A>(L1: Lang, L2: Lang): Lang {
  Alpha(L1.eps || L2.eps, (a: A) => Plus(L1.delta(a), L2.delta(a)))
}

function {:abstemious} Comp<A>(L1: Lang, L2: Lang): Lang {
  Alpha(L1.eps && L2.eps,
        (a: A) => Plus(Comp(L1.delta(a), L2), 
                       Comp(if L1.eps then One() else Zero(), L2.delta(a))))
}

function Star<A>(L: Lang): Lang {
  Alpha(true, (a: A) => Comp(L.delta(a), Star(L)))
}
\end{lstlisting} 

Note the use of the equality-supporting type parameter \lstinline[language=dafny, basicstyle=\small\ttfamily]{==} in the definition of \lstinline[language=dafny, basicstyle=\small\ttfamily]{Singleton}, which restricts the use of the function to types \lstinline[language=dafny, basicstyle=\small\ttfamily]{A} that are known to support run-time equality comparisons (all types support equality in static contexts). In this case, the restriction is needed to ensure the well-definedness of the expression \lstinline[language=dafny, basicstyle=\small\ttfamily]{a == b} in the definition of \lstinline[language=dafny, basicstyle=\small\ttfamily]{Singleton}.

Also note that the \lstinline[language=dafny, basicstyle=\small\ttfamily]@{:abstemious}@ attribute above signals that a function does not need to unfold a codatatype
instance very far (perhaps just one destructor call) to prove a relevant property. Knowing this is the case can aid in
proofs of the properties of the function. In this case, it is needed to convince Dafny that the corecursive calls in
\lstinline[language=dafny, basicstyle=\small\ttfamily]{Comp} and \lstinline[language=dafny, basicstyle=\small\ttfamily]{Star} are logically consistent.

In the language of Coalgebra, the above is best described by us equipping \lstinline[language=dafny, basicstyle=\small\ttfamily]{Lang} with an algebra structure for the functor $\Sigma$. To derive a function such as e.g. \lstinline[language=dafny, basicstyle=\small\ttfamily]{Comp} above, one gives the product \lstinline[language=dafny, basicstyle=\small\ttfamily]{(Lang, Lang)} an appropriate $B$-coalgebra structure and deduces a unique morphism \lstinline[language=dafny, basicstyle=\small\ttfamily]{(Lang, Lang) -> Lang} from the finality of \lstinline[language=dafny, basicstyle=\small\ttfamily]{Lang} as $B$-coalgebra \cite{Rutten:00:Universal}.

\subsection{Denotational Semantics as Induced Morphism}

\label{sec:denotational_semantics_as_induced_morphism}

\begin{figure}[t]
\center
\begin{tikzcd}[column sep=0.6in, row sep=0.1in]
 \Sigma(\texttt{Exp}) \arrow{d}{} \arrow[dashed]{r}{\Sigma(\texttt{Denotational})}  & \Sigma(\texttt{Lang}) \arrow{d}{\lbrack \texttt{Zero}, \texttt{One}, \texttt{Singleton}, \texttt{Plus}, \texttt{Comp}, \texttt{Star} \rbrack} \\
\texttt{Exp} \arrow[dashed]{r}{\texttt{Denotational}}  & \texttt{Lang}
\end{tikzcd}
\begin{tikzcd}[column sep=0.6in, row sep=0.1in]
 \texttt{Exp} \arrow{d}{\langle \texttt{Eps}, \texttt{Delta} \rangle} \arrow[dashed]{r}{\texttt{Operational}}  & \texttt{Lang} \arrow{d}{} \\
B(\texttt{Exp}) \arrow[dashed]{r}{B(\texttt{Operational})}  & B(\texttt{Lang})
\end{tikzcd}
\caption{\lstinline[language=dafny, basicstyle=\small\ttfamily]{Denotational} and \lstinline[language=dafny, basicstyle=\small\ttfamily]{Operational} as induced unique $\Sigma$-algebra and $B$-coalgebra homomorphisms, respectively.}
\label{fig:induced_morphisms}	
\end{figure}

The denotational semantics of regular expressions can now be defined through induction, as a function
\lstinline[language=dafny, basicstyle=\small\ttfamily]{Denotational: Exp -> Lang}, by making use of the operations on languages we have just defined in \Cref{sec:an_algebra_of_formal_languages}. For the sake of clarity, we encapsulate those in a module named  \lstinline[language=dafny, basicstyle=\small\ttfamily]{Languages}:

\begin{lstlisting}[language=dafny]
function Denotational<A(==)>(e: Exp): Lang {
  match e
  case Zero => Languages.Zero()
  case One => Languages.One()
  case Char(a) => Languages.Singleton(a)
  case Plus(e1, e2) => Languages.Plus(Denotational(e1), Denotational(e2))
  case Comp(e1, e2) => Languages.Comp(Denotational(e1), Denotational(e2))
  case Star(e1) => Languages.Star(Denotational(e1))
}
\end{lstlisting} 

The high-level view through the lenses of Coalgebra is depicted in \Cref{fig:induced_morphisms}. By the initiality of  \lstinline[language=dafny, basicstyle=\small\ttfamily]{Exp} as algebra for $\Sigma$, there exists a unique morphism \lstinline[language=dafny, basicstyle=\small\ttfamily]{Denotational: Exp -> Lang} that commutes with the algebraic structures (we formally prove the latter in Dafny in \Cref{sec:denotational_semantics_as_algebra_homomorphism}). 

\subsection{Bisimilarity and Coinduction}

\label{sec:bisimilarity_and_coinduction}

Let us briefly recall the notion of bisimilarity of formal languages. A binary relation
\lstinline[language=dafny, basicstyle=\small\ttfamily]{R} between languages is called \emph{bisimulation}, if for any two languages \lstinline[language=dafny, basicstyle=\small\ttfamily]{L1}, \lstinline[language=dafny, basicstyle=\small\ttfamily]{L2} related by \lstinline[language=dafny, basicstyle=\small\ttfamily]{R} the following
holds: i) \lstinline[language=dafny, basicstyle=\small\ttfamily]{L1} contains the empty word iff \lstinline[language=dafny, basicstyle=\small\ttfamily]{L2} does; and ii) for any \lstinline[language=dafny, basicstyle=\small\ttfamily]{a: A}, the derivatives \lstinline[language=dafny, basicstyle=\small\ttfamily]{L1.delta(a)} and
\lstinline[language=dafny, basicstyle=\small\ttfamily]{L2.delta(a)} are again related by \lstinline[language=dafny, basicstyle=\small\ttfamily]{R}. As it turns out, the union of two bisimulations is again a bisimulation. In
consequence, one can combine all possible bisimulations into a single relation: the \emph{greatest} bisimulation. Two languages are called bisimilar if they are related by this greatest bisimulation. In Dafny,
we can formalise the latter as follows:

\begin{lstlisting}[language=dafny]
greatest predicate Bisimilar<A(!new)>[nat](L1: Lang, L2: Lang) {
  && (L1.eps == L2.eps)
  && (forall a :: Bisimilar(L1.delta(a), L2.delta(a)))
}
\end{lstlisting} 

Note that we used Dafny's type-parameter mode \lstinline[language=dafny, basicstyle=\small\ttfamily]{!new}, which restricts the use of \lstinline[language=dafny, basicstyle=\small\ttfamily]{Bisimilar} to values of type \lstinline[language=dafny, basicstyle=\small\ttfamily]{A} that are not heap-based. This is necessary since a \lstinline[language=dafny, basicstyle=\small\ttfamily]{forall} expression involved in a \lstinline[language=dafny, basicstyle=\small\ttfamily]{greatest predicate} definition is not allowed to depend on the set of allocated references.

Two languages that are equal are also bisimilar, but the reverse is not necessarily true, since there is no extensional equality for functions in Dafny.

It is instructive to think of a \lstinline[language=dafny, basicstyle=\small\ttfamily]{greatest predicate} as pure syntactic sugar. Indeed, under the hood, Dafny's
compiler uses the body above to implicitly generate i) for any \lstinline[language=dafny, basicstyle=\small\ttfamily]{k: nat}, a \emph{prefix predicate}  \lstinline[language=dafny, basicstyle=\small\ttfamily]{Bisimilar#[k](L1, L2)} that signifies that the languages  \lstinline[language=dafny, basicstyle=\small\ttfamily]{L1} and  \lstinline[language=dafny, basicstyle=\small\ttfamily]{L2} concur on the first  \lstinline[language=dafny, basicstyle=\small\ttfamily]{k}-unrollings of the definition above; and
ii) a predicate  \lstinline[language=dafny, basicstyle=\small\ttfamily]{Bisimilar(L1, L2)} that is true iff  \lstinline[language=dafny, basicstyle=\small\ttfamily]{Bisimilar#[k](L1, L2)} is true for all  \lstinline[language=dafny, basicstyle=\small\ttfamily]{k: nat}:

\begin{lstlisting}[language=dafny]
/* Pseudo code for illustration purposes */

predicate Bisimilar#<A(!new)>[k: nat](L1: Lang, L2: Lang)
  decreases k
{
  if k == 0 then
    true
  else
    && (L1.eps == L2.eps)
    && (forall a :: Bisimilar#[k-1](L1.delta(a), L2.delta(a)))
}

predicate Bisimilar<A(!new)>(L1: Lang, L2: Lang) {
  forall k: nat :: Bisimilar#[k](L1, L2)
}
\end{lstlisting} 

Note the use of the \lstinline[language=dafny, basicstyle=\small\ttfamily]{decreases} clause in the definition of \lstinline[language=dafny, basicstyle=\small\ttfamily]{Bisimilar#[k](L1, L2)}. Dafny requires us to convince it that all functions terminate. A \lstinline[language=dafny, basicstyle=\small\ttfamily]{decreases} clause is used to support the proof of termination of a function in the presence of recursion.  At each recursive call to a function, Dafny checks that the \lstinline[language=dafny, basicstyle=\small\ttfamily]{decreases} clause is strictly smaller than the one of its caller with respect to a built-in well-founded order. In this case, Dafny verifies the inequality \lstinline[language=dafny, basicstyle=\small\ttfamily]{k-1 < k} with respect to the natural well-founded order \lstinline[language=dafny, basicstyle=\small\ttfamily]{<} of \lstinline[language=dafny, basicstyle=\small\ttfamily]{nat}.

Now that we have its definition in place, let us establish a property about bisimilarity, say, that it is a reflexive
relation. With the \lstinline[language=dafny, basicstyle=\small\ttfamily]{greatest lemma} construct, Dafny is able to derive a proof completely on its own:

\begin{lstlisting}[language=dafny]
greatest lemma BisimilarityIsReflexive<A(!new)>[nat](L: Lang)
  ensures Bisimilar(L, L)
{}
\end{lstlisting} 

Once again, it is instructive to think of a \lstinline[language=dafny, basicstyle=\small\ttfamily]{greatest lemma} as pure syntactic sugar. Under the hood, Dafny's
compiler uses the body of \lstinline[language=dafny, basicstyle=\small\ttfamily]{BisimilarityIsReflexive} above to generate i) for any \lstinline[language=dafny, basicstyle=\small\ttfamily]{k: nat}, a \emph{prefix lemma}
\lstinline[language=dafny, basicstyle=\small\ttfamily]{BisimilarityIsReflexive#[k](L)} that ensures that the prefix predicate \lstinline[language=dafny, basicstyle=\small\ttfamily]{Bisimilar#[k](L, L)} is satisfied; and ii) a
lemma \lstinline[language=dafny, basicstyle=\small\ttfamily]{BisimilarityIsReflexive(L)} that ensures that \lstinline[language=dafny, basicstyle=\small\ttfamily]{Bisimilar(L, L)} is true by calling \lstinline[language=dafny, basicstyle=\small\ttfamily]{BisimilarityIsReflexive#[k](L, L)}
for all \lstinline[language=dafny, basicstyle=\small\ttfamily]{k: nat}:\newpage

\begin{lstlisting}[language=dafny]
/* Pseudo code for illustration purposes */

lemma BisimilarityIsReflexive#<A(!new)>[k: nat](L: Lang)
  ensures Bisimilar#[k](L, L)
  decreases k
{
  if k == 0 {
  } else {
    forall a ensures Bisimilar#[k-1](L.delta(a), L.delta(a)) {
      BisimilarityIsReflexive#[k-1](L.delta(a));
    }
  }
} 

lemma BisimilarityIsReflexive<A(!new)>(L: Lang)
  ensures Bisimilar(L, L)
{
  forall k: nat ensures Bisimilar#[k](L, L) {
    BisimilarityIsReflexive#[k](L);
  }
}
\end{lstlisting} 

We refer the reader interested in further details about Dafny's take on coinduction, predicates, and ordinals to \cite{Leino:coinduction}.

\subsection{Denotational Semantics as Algebra Homomorphism}

\label{sec:denotational_semantics_as_algebra_homomorphism}

In this section, we are interested in homomorphisms of type \lstinline[language=dafny, basicstyle=\small\ttfamily]{f: Exp -> Lang} (more precisely, in \lstinline[language=dafny, basicstyle=\small\ttfamily]{Denotational}), that is, functions which commute, up to bisimilarity, with the algebra structures we encountered in \Cref{sec:regular_expressions_as_datatype} and \Cref{sec:an_algebra_of_formal_languages}, respectively. In Dafny, we call such functions simply algebra homomorphisms. We define pointwise commutativity by comparing languages for bisimilarity:

\begin{lstlisting}[language=dafny]
ghost predicate IsAlgebraHomomorphism<A(!new)>(f: Exp -> Lang) {
  forall e :: IsAlgebraHomomorphismPointwise(f, e)
}

ghost predicate IsAlgebraHomomorphismPointwise<A(!new)>
  (f: Exp -> Lang, e: Exp) {
  Bisimilar<A>(
    f(e),
    match e
    case Zero => Languages.Zero()
    case One => Languages.One()
    case Char(a) => Languages.Singleton(a)
    case Plus(e1, e2) => Languages.Plus(f(e1), f(e2))
    case Comp(e1, e2) => Languages.Comp(f(e1), f(e2))
    case Star(e1) => Languages.Star(f(e1))
  )
}
\end{lstlisting} 

Note that we used the \lstinline[language=dafny, basicstyle=\small\ttfamily]{ghost} modifier (which signals that an entity is meant for specification only, not for compilation). A \lstinline[language=dafny, basicstyle=\small\ttfamily]{greatest predicate} is always implicitly \lstinline[language=dafny, basicstyle=\small\ttfamily]{ghost}, so \lstinline[language=dafny, basicstyle=\small\ttfamily]{IsAlgebraHomomorphismPointwise} must be declared \lstinline[language=dafny, basicstyle=\small\ttfamily]{ghost} to call \lstinline[language=dafny, basicstyle=\small\ttfamily]{Bisimilar}, and \lstinline[language=dafny, basicstyle=\small\ttfamily]{IsAlgebraHomomorphism} must be declared \lstinline[language=dafny, basicstyle=\small\ttfamily]{ghost} to call \lstinline[language=dafny, basicstyle=\small\ttfamily]{IsAlgebraHomomorphismPointwise}.

The proof that \lstinline[language=dafny, basicstyle=\small\ttfamily]{Denotational} is an algebra homomorphism is straightforward; it essentially follows from bisimilarity being reflexive:
\newpage
\begin{lstlisting}[language=dafny]
lemma DenotationalIsAlgebraHomomorphism<A(!new)>()
  ensures IsAlgebraHomomorphism<A>(Denotational)
{
  forall e ensures IsAlgebraHomomorphismPointwise<A>(Denotational, e) {
    BisimilarityIsReflexive<A>(Denotational(e));
  }
}
\end{lstlisting} 

\vspace{-0.5cm}

\section{Operational Semantics}

\label{sec:operational_semantics}

In this section, we provide an alternative perspective on the semantics of regular expressions. In Dafny, we equip the set of regular expressions with a coalgebraic structure of the type of unpointed deterministic automata, formalise its \emph{operational} semantics as a function from regular expressions to formal languages, and prove that the latter is a coalgebra homomorphism.

\subsection{A Coalgebra of Regular Expressions}

\label{sec:a_coalgebra_of_regular_expressions}

In \Cref{sec:an_algebra_of_formal_languages} we equipped the set of formal languages with an algebraic structure that resembled the one of regular expressions. Now, we are aiming for the dual: we would like to equip the set of regular expressions with a coalgebraic structure that resembles the one of formal languages. More concretely, we would like to turn the set of regular expressions into a $B$-coalgebra, that is, a deterministic automaton (without initial state) in which a state \lstinline[language=dafny, basicstyle=\small\ttfamily]{e} is i) accepting iff \lstinline[language=dafny, basicstyle=\small\ttfamily]{Eps(e) == true} and ii) transitions to a state \lstinline[language=dafny, basicstyle=\small\ttfamily]{Delta(e)(a)} if given the input \lstinline[language=dafny, basicstyle=\small\ttfamily]{a: A}. Note how our definitions resemble the Brzozowski derivatives:

\begin{lstlisting}[language=dafny]
function Eps<A>(e: Exp): bool {
  match e
  case Zero => false
  case One => true
  case Char(a) => false
  case Plus(e1, e2) => Eps(e1) || Eps(e2)
  case Comp(e1, e2) => Eps(e1) && Eps(e2)
  case Star(e1) => true
}

function Delta<A(==)>(e: Exp): A -> Exp {
  (a: A) =>
    match e
    case Zero => Zero
    case One => Zero
    case Char(b) => if a == b then One else Zero
    case Plus(e1, e2) => Plus(Delta(e1)(a), Delta(e2)(a))
    case Comp(e1, e2) => 
      Plus(Comp(Delta(e1)(a), e2), Comp(if Eps(e1) then One else Zero, Delta(e2)(a)))
    case Star(e1) => Comp(Delta(e1)(a), Star(e1))
}
\end{lstlisting} 

\subsection{Operational Semantics as Induced Morphism}

\label{sec:operational_semantics_as_induced_morphism}

The operational semantics of regular expressions can now in Dafny be defined via coinduction, as a function \lstinline[language=dafny, basicstyle=\small\ttfamily]{Operational: Exp -> Lang}, by making use of the coalgebraic structure on expressions for the functor $B$ we have just defined in \Cref{sec:a_coalgebra_of_regular_expressions}:

\begin{lstlisting}[language=dafny]
function Operational<A(==)>(e: Exp): Lang {
  Alpha(Eps(e), (a: A) => Operational(Delta(e)(a)))
}
\end{lstlisting} 

The high-level view through the lenses of Coalgebra is depicted in \Cref{fig:induced_morphisms}. By the finality of  \lstinline[language=dafny, basicstyle=\small\ttfamily]{Lang} as $B$-coalgebra, there exists a unique morphism \lstinline[language=dafny, basicstyle=\small\ttfamily]{Operational: Exp -> Lang} that commutes with the $B$-coalgebra structures (the latter is formally proven in Dafny in \Cref{sec:operational_semantics_as_coalgebra_homomorphism}).

\subsection{Operational Semantics as Coalgebra Homomorphism}

\label{sec:operational_semantics_as_coalgebra_homomorphism}

In \Cref{sec:denotational_semantics_as_algebra_homomorphism} we defined in Dafny algebra homomorphisms as functions of type \lstinline[language=dafny, basicstyle=\small\ttfamily]{Exp -> Lang} that commute, up to bisimilarity, with the $\Sigma$-algebra structures of regular expressions and formal languages, respectively. Analogously, we now define a function of the same type as coalgebra homomorphism, if it commutes, up to pointwise bisimilarity, with the $B$-coalgebra structures of regular expressions and formal languages, respectively:

\begin{lstlisting}[language=dafny]
ghost predicate IsCoalgebraHomomorphism<A(!new)>(f: Exp -> Lang) {
  && (forall e :: f(e).eps == Eps(e))
  && (forall e, a :: Bisimilar(f(e).delta(a), f(Delta(e)(a))))
}
\end{lstlisting} 

It is straightforward to formally prove that \lstinline[language=dafny, basicstyle=\small\ttfamily]{Operational} is a coalgebra homomorphism in the above sense: once again, the central argument is that bisimilarity is a reflexive relation.

\begin{lstlisting}[language=dafny]
lemma OperationalIsCoalgebraHomomorphism<A(!new)>()
  ensures IsCoalgebraHomomorphism<A>(Operational)
{
  forall e, a ensures Bisimilar<A>(Operational(e).delta(a), Operational(Delta(e)(a))) {
    BisimilarityIsReflexive(Operational(e).delta(a));
  }
}
\end{lstlisting} 

\vspace{-0.42cm}

\section{Well-Behaved Semantics}

\label{sec:well-behaved_semantics}

So far, we have seen two dual approaches for assigning formal language semantics to regular expressions:

\begin{itemize}
	\item \lstinline[language=dafny, basicstyle=\small\ttfamily]{Denotational}: an algebra homomorphism obtained via induction
	\item \lstinline[language=dafny, basicstyle=\small\ttfamily]{Operational}: a coalgebra homomorphism obtained via coinduction
\end{itemize}

Next, we show in Dafny that the denotational and operational semantics of regular expressions are \emph{well-behaved} (a term we adapt from \cite{Turi:1997:Mathematical}): they constitute two sides of the same coin. First, we show that \lstinline[language=dafny, basicstyle=\small\ttfamily]{Denotational} is also a coalgebra homomorphism, and that coalgebra homomorphisms are unique up to bisimilarity. We then deduce from the former that \lstinline[language=dafny, basicstyle=\small\ttfamily]{Denotational} and \lstinline[language=dafny, basicstyle=\small\ttfamily]{Operational} coincide pointwise, up to bisimilarity. Finally, we show that \lstinline[language=dafny, basicstyle=\small\ttfamily]{Operational} is also an algebra homomorphism.

\subsection{Denotational Semantics as Coalgebra Homomorphism}

\label{sec:denotational_semantics_as_coalgebra_homomorphism}

In this section, we establish that \lstinline[language=dafny, basicstyle=\small\ttfamily]{Denotational} not only commutes with the algebraic structures of regular expressions and formal languages but also with their coalgebraic structures:

\begin{lstlisting}[language=dafny]
lemma DenotationalIsCoalgebraHomomorphism<A(!new)>()
  ensures IsCoalgebraHomomorphism<A>(Denotational)
\end{lstlisting} 

The proof of the lemma is a bit more elaborate than the ones we have encountered so far. It can be divided into two subproofs, both of which make use of induction. One of the subproofs is straightforward, the other, more difficult one, again uses the reflexivity of bisimilarity, but also that the latter is a congruence relation with respect to \lstinline[language=dafny, basicstyle=\small\ttfamily]{Plus} and \lstinline[language=dafny, basicstyle=\small\ttfamily]{Comp}:

\begin{lstlisting}[language=dafny]
greatest lemma PlusCongruence<A(!new)>[nat]
  (L1a: Lang, L1b: Lang, L2a: Lang, L2b: Lang)
  requires Bisimilar(L1a, L1b)
  requires Bisimilar(L2a, L2b)
  ensures Bisimilar(Plus(L1a, L2a), Plus(L1b, L2b))
{}

lemma CompCongruence<A(!new)>(L1a: Lang, L1b: Lang, L2a: Lang, L2b: Lang)
  requires Bisimilar(L1a, L1b)
  requires Bisimilar(L2a, L2b)
  ensures Bisimilar(Comp(L1a, L2a), Comp(L1b, L2b))
\end{lstlisting} 

Dafny is able to prove  \lstinline[language=dafny, basicstyle=\small\ttfamily]{PlusCongruence} on its own, as it can take advantage of the syntactic sugaring of the  \lstinline[language=dafny, basicstyle=\small\ttfamily]{greatest lemma} construct. For  \lstinline[language=dafny, basicstyle=\small\ttfamily]{CompCongruence} we have to put in a bit of manual work ourselves.

\subsection{Coalgebra Homomorphisms Are Unique}

\label{sec:coalgebra_homomorphisms_are_unique}

The aim of this section is to show in Dafny that, up to pointwise bisimilarity, there only exists one coalgebra homomorphism of type \lstinline[language=dafny, basicstyle=\small\ttfamily]{Exp -> Lang}:

\begin{lstlisting}[language=dafny]
lemma UniqueCoalgebraHomomorphism<A(!new)>(f: Exp -> Lang, g: Exp -> Lang, e: Exp)
  requires IsCoalgebraHomomorphism(f)
  requires IsCoalgebraHomomorphism(g)
  ensures Bisimilar(f(e), g(e))
\end{lstlisting} 

Of course, the perspective of Coalgebra suggests that the statement may in fact be strengthened to: for \emph{any} coalgebra  \lstinline[language=dafny, basicstyle=\small\ttfamily]{C} there exists exactly one coalgebra homomorphism of type  \lstinline[language=dafny, basicstyle=\small\ttfamily]{C -> Lang}, up to pointwise bisimilarity. For our purposes, the weaker statement above will be sufficient. At the heart of the proof lies the observation that bisimilarity is transitive:

\begin{lstlisting}[language=dafny]
greatest lemma BisimilarityIsTransitive<A(!new)>[nat](L1: Lang, L2: Lang, L3: Lang)
  requires Bisimilar(L1, L2) && Bisimilar(L2, L3)
  ensures Bisimilar(L1, L3)
{}
\end{lstlisting} 

In fact, in practice, we actually use a slightly more fine-grained formalisation, as is illustrated below by the call to \lstinline[language=dafny, basicstyle=\small\ttfamily]{BisimilarityIsTransitivePointwise} in the proof of \lstinline[language=dafny, basicstyle=\small\ttfamily]{UniqueCoalgebraHomomorphismHelperPointwise}, which in turn is used to prove \lstinline[language=dafny, basicstyle=\small\ttfamily]{UniqueCoalgebraHomomorphism}:

\begin{lstlisting}[language=dafny]
lemma UniqueCoalgebraHomomorphismHelperPointwise<A(!new)>
  (k: nat, f: Exp -> Lang, g: Exp -> Lang, L1: Lang, L2: Lang)
  requires IsCoalgebraHomomorphism(f)
  requires IsCoalgebraHomomorphism(g)
  requires exists e :: Bisimilar#[k](L1, f(e)) && Bisimilar#[k](L2, g(e))
  ensures Bisimilar#[k](L1, L2)
{
  var e :| Bisimilar#[k](L1, f(e)) && Bisimilar#[k](L2, g(e));
  if k != 0 {
    forall a ensures Bisimilar#[k-1](L1.delta(a), L2.delta(a)) {
      BisimilarityIsTransitivePointwise(
        k-1, L1.delta(a),  f(e).delta(a), f(Delta(e)(a))
      );
      BisimilarityIsTransitivePointwise(
        k-1, L2.delta(a),  g(e).delta(a), g(Delta(e)(a))
      );
      UniqueCoalgebraHomomorphismHelperPointwise(
        k-1, f, g, L1.delta(a), L2.delta(a)
      );
    }
}}

lemma BisimilarityIsTransitivePointwise<A(!new)>(k: nat, L1: Lang, L2: Lang, L3: Lang)
  ensures Bisimilar#[k](L1, L2) && Bisimilar#[k](L2, L3) ==> Bisimilar#[k](L1, L3)
{
  if k != 0 {
    if Bisimilar#[k](L1, L2) && Bisimilar#[k](L2, L3) {
      assert Bisimilar#[k](L1, L3) by {
        forall a ensures Bisimilar#[k-1](L1.delta(a), L3.delta(a)) {
          BisimilarityIsTransitivePointwise(k-1, L1.delta(a), L2.delta(a), L3.delta(a));
        }
}}}}
\end{lstlisting} 

Note the use of Dafny's let-such-that assignment \lstinline[language=dafny, basicstyle=\small\ttfamily]{:|} in the body of the lemma \lstinline[language=dafny, basicstyle=\small\ttfamily]{UniqueCoalgebraHomomorphismHelperPointwise}. For any predicate \lstinline[language=dafny, basicstyle=\small\ttfamily]{P}, the expression \lstinline[language=dafny, basicstyle=\small\ttfamily]{x :| P(x)} assigns a value to \lstinline[language=dafny, basicstyle=\small\ttfamily]{x} such that \lstinline[language=dafny, basicstyle=\small\ttfamily]{P(x)} is true. The predicate \lstinline[language=dafny, basicstyle=\small\ttfamily]{P} needs to be non-empty, but in a ghost context doesn't have to constrain \lstinline[language=dafny, basicstyle=\small\ttfamily]{x} uniquely: the choice of the latter is non-deterministic. To be compilable, the value of a let-such-that expression must be uniquely determined, however. In this case, the precondition of the lemma guarantees that the predicate is non-empty. For further background, we refer the reader to the Dafny Power User note \cite{Leino:iteratingovercollection}.

\subsection{Denotational and Operational Semantics Are Bisimilar}

\label{sec:denotational_and_operational_semantics_are_bisimilar}

From the previous results, we can immediately deduce our main claim that denotational and operational semantics coincide, up to pointwise bisimilarity:

\begin{lstlisting}[language=dafny]
lemma OperationalAndDenotationalAreBisimilar<A(!new)>(e: Exp)
  ensures Bisimilar<A>(Operational(e), Denotational(e))
{
  OperationalIsCoalgebraHomomorphism<A>();
  DenotationalIsCoalgebraHomomorphism<A>();
  UniqueCoalgebraHomomorphism<A>(Operational, Denotational, e);
}
\end{lstlisting} 

\subsection{Operational Semantics as Algebra Homomorphism}

\label{sec:operational_semantics_as_algebra_homomorphism}

\begin{figure}[t]
\center
\begin{tikzcd}[column sep=2in, row sep=0.3cm]
 \Sigma(\texttt{Exp}) \arrow{d}{} \arrow[dashed]{r}{\Sigma(\texttt{Denotational} \cong \texttt{Denotational})}  & \Sigma(\texttt{Lang}) \arrow{d}{\lbrack \texttt{Zero}, \texttt{One}, \texttt{Singleton}, \texttt{Star}, \texttt{Plus}, \texttt{Comp} \rbrack} \\
\texttt{Exp} \arrow[dashed]{r}{\texttt{Denotational} \cong \texttt{Operational}}  \arrow{d}[left]{\langle \texttt{Eps}, \texttt{Delta} \rangle}  & \texttt{Lang} \arrow{d}{} \\
B(\texttt{Exp}) \arrow[dashed]{r}{B( \texttt{Denotational} \cong \texttt{Operational})}  & B(\texttt{Lang})
\end{tikzcd}
\caption{The \lstinline[language=dafny, basicstyle=\small\ttfamily]{Denotational} and \lstinline[language=dafny, basicstyle=\small\ttfamily]{Operational} semantics of regular expressions are well-behaved.}
\label{fig:well_behaved}	
\end{figure}

As a bonus, for the sake of symmetry, let us also prove that \lstinline[language=dafny, basicstyle=\small\ttfamily]{Operational} is an algebra homomorphism. (We already know that it is a coalgebra homomorphism, and that \lstinline[language=dafny, basicstyle=\small\ttfamily]{Denotational} is both an algebra and coalgebra homomorphism.)

\begin{lstlisting}[language=dafny]
lemma OperationalIsAlgebraHomomorphism<A(!new)>()
  ensures IsAlgebraHomomorphism<A>(Operational)
\end{lstlisting} 

The idea of the proof is to take advantage of \lstinline[language=dafny, basicstyle=\small\ttfamily]{Denotational} being an algebra homomorphism, by translating its properties to \lstinline[language=dafny, basicstyle=\small\ttfamily]{Operational} via the lemma in \Cref{sec:denotational_and_operational_semantics_are_bisimilar}. The relevant new statements capture that bisimilarity is symmetric and a congruence with respect to the \lstinline[language=dafny, basicstyle=\small\ttfamily]{Star} operation:

\begin{lstlisting}[language=dafny]
  greatest lemma BisimilarityIsSymmetric<A(!new)>[nat](L1: Lang, L2: Lang)
    ensures Bisimilar(L1, L2) ==> Bisimilar(L2, L1)
    ensures Bisimilar(L1, L2) <== Bisimilar(L2, L1)
  {}

  lemma StarCongruence<A(!new)>(L1: Lang, L2: Lang)
    requires Bisimilar(L1, L2)
    ensures Bisimilar(Star(L1), Star(L2))
\end{lstlisting} 

The full picture is depicted in \Cref{fig:well_behaved}: \lstinline[language=dafny, basicstyle=\small\ttfamily]{Denotational} is induced by the initiality of \lstinline[language=dafny, basicstyle=\small\ttfamily]{Exp} and \lstinline[language=dafny, basicstyle=\small\ttfamily]{Operational} is induced by the finality of \lstinline[language=dafny, basicstyle=\small\ttfamily]{Lang}. By uniqueness, the two homomorphisms coincide, up to pointwise bisimilarity -- the semantics of regular expressions are well-behaved.

\section{Discussion and Related Work}

\label{sec:discussion_and_related_work}
We have used Dafny's built-in inductive and coinductive reasoning capabilities to define the denotational and operational semantics of regular expressions and to prove that they are well-behaved. The concept of well-behaved semantics, in the context of bialgebras (which consist of an algebra and a coalgebra over the same carrier that interact with each other in a suitable way), goes back to Turi and Plotkin~\cite{Turi:1997:Mathematical} and was adapted by Jacobs to the case of regular expressions \cite{Jacobs:2006:Bialgebraic}. The bialgebraic perspective on regular expressions can be thought of as a generalisation of the classical automata-theoretic construction from Brzozowski in the 1960s \cite{brzozowski1964derivatives}. A more modern presentation can be found in e.g.~\cite{Silva:2010:Kleene}. The coalgebraic aspects build on results by Rutten~\cite{Rutten:00:Universal}, Gumm~\cite{Gumm:2000:Elements}, and others~\cite{jacobs2012trace}. As our presentation focused on the most important and high-level aspects of the proofs in Dafny, we invite the interested reader to take a look at the full Dafny source code \cite{Zetzsche:2024:Well-behaved}.

The work closest to ours in spirit is \cite{traytel2017formal}, in which the authors use Isabelle~\cite{Paulson:1988:Isabelle}, an LCF-style interactive theorem prover, to prove that formal languages represented as a coinductively defined trie satisfy the axioms of \emph{Kleene Algebra} (KA) \cite{Kozen:1994:Completeness}. As for the differences, the authors of \cite{traytel2017formal} don't touch on the aspect of well-behaved semantics, and we leave a formal proof of the axioms of KA in Dafny as future work. For the latter, because of the interplay between algebraic and coalgebraic structures, we plan to employ so-called up-to-techniques~\cite{Rot:2013:Coinductive}, which allow for compact coinductive proofs by making use of the underlying algebraic structure. Further related to this paper and \cite{traytel2017formal} are \cite{blanchette2017friends,blanchette2015foundational}, in which the implementation of corecursion in Isabelle is discussed.

We depart from other work~\cite{moreira2015deciding,Kraus:2012:Regular,Paulson:2015:Formalisation}, which models formal languages intrinsically, as sets of words, and consequently begins by equipping the set of languages with an appropriate coalgebraic structure, whereas our extrinsic treatment in Dafny essentially axiomatises the latter.

\section{Future Work}
\label{sec:future_work}

Besides proving in Dafny the soundness of Kleene Algebra axioms for coalgebraically defined languages, we are mainly interested in adapting our formalisation of the semantics of regular expressions and the proof of their well-behavedness to other theories. 

An immediate target is \emph{Kleene Algebra with Tests} (KAT) \cite{Kozen:1997:KAT}, which extends the theory of regular expressions, Kleene Algebra, with so-called \emph{tests} (elements of a finitely generated Boolean Algebra). KAT can be used to reason about the equivalence of uninterpreted imperative programs, with tests used to model program guards. The theory has been successfully applied to program schematology~\cite{Angus:2001:Kleene} and has been used to reason about compiler optimizations~\cite{Kozen:2000:Certification} and cache control~\cite{Barth:2002:Equational}. KAT admits both denotational semantics, through so-called guarded string languages, and operational semantics, through so-called automata on guarded strings \cite{Kozen:2001:Automata}. 

There are more natural targets since KAT has been further extended in multiple directions. One such example is \emph{Guarded Kleene Algebra with Tests} (GKAT) \cite{Smolka:2020:Guarded}, an efficiently decidable fragment of KAT that admits operational semantics through strictly deterministic automata on guarded strings~\cite{Smolka:2020:Guarded}. Another example is \emph{NetKAT} ~\cite{Anderson:2014:NetKAT}, which extends KAT with primitives that allow the reasoning about Software Defined Networks. The verification of properties of such networks can be reduced to deciding the equivalence of NetKAT expressions, which in turn relies on their operational semantics \cite{Foster:2015:Coalgebraic}. 

Overall, we hope that the present formalisation both illustrates Dafny's potential for coalgebraic reasoning and serves as a blueprint for further adaption. 
\begin{credits}
\subsubsection{\ackname} 
The authors are thankful to Aaron Tomb and Rustan Leino for their comments on an earlier version of this paper. 
\subsubsection{\discintname}
The authors have no competing interests to declare that are
relevant to the content of this article.
\end{credits}

 \bibliographystyle{splncs04}
 \bibliography{bibliography}

\end{document}